\documentclass[journal=enfuem]{achemso}

\usepackage{xr}
\usepackage{comment}
\usepackage{cuted}

\usepackage[T1]{fontenc}

\usepackage{geometry}
\usepackage{setspace}

\usepackage{achemso}

\usepackage{graphicx}
\usepackage{float}
\newfloat{scheme}{htbp}{los}
\floatname{scheme}{Scheme}
\floatname{chart}{Chart}
\newfloat{graph}{htbp}{loh}
\usepackage[colorlinks=true, urlcolor=blue]{hyperref}
\usepackage{chemformula} 
\usepackage[version = 4]{mhchem} 
\usepackage{siunitx}
\DeclareSIUnit\angstrom{\text{\AA}}
\DeclareSIUnit[number-unit-product = {\,}]\cal{cal}
\DeclareSIUnit\kcal{\kilo\cal}

\author{Felipe Silva Carvalho}
\email{fscarv@caltech.edu}
\affiliation{Division of Chemistry and Chemical Engineering, California Institute of Technology, Pasadena, CA 91125, USA}

\author{Tyler Luchko}
\affiliation{Department of Physics and Astronomy, California State University, Northridge, CA 91325, USA}

\author{Zhen-Gang Wang}
\affiliation{Division of Chemistry and Chemical Engineering, California Institute of Technology, Pasadena, CA 91125, USA}

\title{Methane and ethane adsorption on calcite surfaces: effects of surface morphology from molecular dynamics and 3D-RISM}

\begin{document}

\maketitle

\begin{abstract}
Understanding the adsorption of alkanes on mineral surfaces is important for industrial processes such as oil and gas recovery from shale reservoirs. Previous studies have examined the adsorption of functionalized and non-functionalized alkanes on calcite, as well as adsorption under confinement. However, real surfaces commonly exhibit defects that can influence adsorption behavior. In this work, molecular dynamics (MD) simulations and the three-dimensional reference interaction site model (3D-RISM) are used to investigate the adsorption of methane and ethane on calcite surfaces with step defects. MD simulations were carried out with and without positional restraints to investigate the effects of carbonate reorientation. 3D-RISM calculations were performed on these structures and on energy-minimized configurations of the relaxed crystal. Both methods show adsorption enhanced at step defects relative to flat regions, with methane exhibiting higher excess adsorption than ethane. Carbonate reorientation at the relaxed step edges reduces the local density peaks at specific step sites in both methods, while the integrated excess adsorption from MD remains comparable to that of the restrained surface. 3D-RISM predicts lower maxima for the highest density peaks and larger excess adsorption than MD, but reproduces the spatial organization and overall adsorption trends at a fraction of the computational cost. These results highlight the importance of surface heterogeneity in fluid-mineral interactions and demonstrate that 3D-RISM is an efficient tool for semi-quantitative screening of adsorption across complex surface geometries
\end{abstract}

\vskip 0.2in
Keywords: Alkanes; 3D-RISM; Molecular dynamics; Calcite; Step defects.

\section{Introduction}
Calcite is one of the most common minerals, making up about 4\% of Earth's crust.\cite{heberling2011structure} It plays an important role in natural and industrial processes, such as biomineralization\cite{arias2008polysaccharides}, CO\textsubscript{2} geosequestration\cite{huijgen2005mineral}, and enhanced oil recovery\cite{kumar2025enhanced}.

Several studies have investigated adsorption of organic molecules on calcite surfaces, including functionalized $n$-alkanes such as hexanoic acid\cite{ghatee2015molecular}, nonanoic acid\cite{sedghi2016atomistic}, octadecanoic acid\cite{duffy2002modelling}, and polystyrene sulfonate\cite{shen2013understanding}. Adsorption of non-functionalized alkanes, such as $n$-hexane and $n$-dodecane, has also been studied,\cite{li2021modeling} reporting the formation of distinct adsorption layers. These results suggest that the formation of ordered adsorption structures on calcite does not necessarily require specific functional groups. Additionally, structural and dynamical properties of CO\textsubscript{2} and $n$-alkane mixtures confined in nanopores have been investigated.\cite{santos2018molecular,santos2022molecular}

However, these studies typically consider ideal flat calcite surfaces. Experimental observations by Stipp \textit{et al.}\cite{stipp1994calcite} showed that freshly fractured calcite surfaces exhibit terraces with step edges, and that exposure to air or water leads to the formation of pits and irregular features due to differential dissolution.

Some studies have addressed molecule adsorption on stepped calcite surfaces, particularly in aqueous environments. For instance, differences in adsorption between flat and stepped surfaces have been reported for calcium-uranyl-carbonate complexes\cite{doudou2012atomistic}, while step edges were identified as preferential binding sites for phosphonates\cite{chen2013molecular} and other organic species and ions\cite{aufort2022determining}. However, analysis of step defects effect on molecule-surface interactions for simple hydrocarbons remain scarce, and a systematic investigation of the influence of step defects on the adsorption behavior of light alkanes has not been done, yet.

In this work, we take a first step towards a systematic investigation of methane and ethane adsorption on flat and stepped calcite surfaces, employing molecular dynamics (MD) simulations and the three-dimensional reference interaction site model\cite{beglov_integral_1997,chandler_optimized_1972,hirata_extended_1981,hirata_three-dimensional_2004,kovalenko_three-dimensional_1998} (3D-RISM). Particular attention is given to the comparison between these approaches and to the role of structural features such as step edges. 

3D-RISM treats the solid surface explicitly while describing the fluid through statistical thermodynamics, enabling the calculation of equilibrium solvent distributions without the sampling limitations typically associated with molecular dynamics simulations. Because the 3D-RISM formalism is solved for a fixed solute configuration, it is particularly well suited for crystalline systems where atoms fluctuate around equilibrium positions. However, this fixed-structure assumption may become critical for defected surfaces, where local structural relaxation can modify adsorption sites and interfacial properties.

To assess the influence of surface relaxation and to enable a meaningful comparison with MD, simulations were therefore performed using two protocols: (i) a fully restrained crystal structure, which preserves the experimental geometry, and (ii) a minimally restrained system in which only a single atom was fixed to prevent rigid-body translation, allowing the stepped surface to relax. 3D-RISM calculations were performed on the unrelaxed structures and also for a set of different surface configurations to account for thermal fluctuations and relaxation. In addition, structural optimization of the defected calcite surface was performed within the 3D-RISM framework to evaluate whether the relaxed configuration observed in MD can be reproduced in this framework.

\section{Methods}
\subsection{RISM theory}
The details of RISM theory have been given elsewhere\cite{kovalenko1999solution,caseAmberTools2023,case2025amber}, here we give a brief description. Before solving the 3D-RISM equations for the solvent in the presence of a solute, it is necessary to acquire the correlation functions for the bulk solvent phase. This is done by solving 1D-RISM equation:\cite{chandler1972optimized}
\begin{equation}
h_{\alpha\gamma}(r) = \sum_{\mu\nu} \int\int \omega_{\alpha\mu}(|\mathbf{r}_1 - \mathbf{r}'|)c_{\mu\nu}(|\mathbf{r}' - \mathbf{r}''|) \times \left[ \omega_{\nu\gamma}(|\mathbf{r}''-\mathbf{r}_2|) + \rho_\nu h_{\nu\gamma}(|\mathbf{r}''-\mathbf{r}_2|) \right]d\mathbf{r}'d\mathbf{r}''
\label{eq:rism1d}
\end{equation}
where $r=|\mathbf{r}_2-\mathbf{r}_1|$, Greek letters in the subscripts denote fluid phase sites, $h(r)$ is the total correlation function, $c(r)$ is the direct corration function, $\rho$ is the bulk density, and $\omega(r)$ is the intramolecular correlation function, which contains structural information of the fluid molecules. Then, distribution of the liquid around the solute may be obtained by solving the 3D-RISM equations, given by\cite{kovalenko1998three}

\begin{equation}
h_\gamma^{UV}(\mathbf{r}) = \sum_\alpha \int c_{\alpha}^{UV}(|\mathbf{r} - \mathbf{r}'|) \times \left[ \omega_{\alpha\gamma}^{VV}(\mathbf{r}') + \rho_\alpha h_{\alpha\gamma}^{VV}(\mathbf{r}') \right]d\mathbf{r}' 
\label{eq:rism3d}
\end{equation}
in which the Greek letter subscript indicates fluid phase sites and the superscripts UV and VV indicate solute-solvent and solvent-solvent correlation functions; the latter are acquired from equation (\ref{eq:rism1d}). As the total and direct correlation functions are unknown, it must be self-consistently solved with a closure relation, which introduces interaction potential energy and temperature. In this work, the partial series expansions of order 1 (PSE-1, also known as Kovalenko-Hirata, KH)\cite{kovalenko1999self} and 2 (PSE-2)\cite{kast2008closed} closure relations will be used for comparing the bulk fluid results and PSE-2 will be used for the solid-fluid interface.

\subsection{Force field}
\subsubsection{Fluid phase}
The transferable potentials for phase equilibria (TraPPE) force field, a united atom model with no partial charges, was used for the alkanes.\cite{martin1998transferable} Since this is not an Amber force field, custom \verb|.frcmod| and \verb|.lib| files were generated for use in \verb|tleap| (see Supporting Information for details).

The \verb|.mdl| files necessary to run 1D-RISM calculations were generated from the topology, \verb|.parm7|, and coordinate, \verb|.rst7|, files for a single molecule of the fluid, which were generated from the \verb|.frcmod|, \verb|.lib| and \verb|.pdb| files using \verb|generateMDL|\cite{silva2026development}, with modifications that will be incorporated in a future release.

\subsubsection{Crystal}
Following protocols established in previous studies,\cite{glisman2024binding,mantha2024adsorption,glisman2024multivalent} the calcite surface was parameterized with the general Amber force field.\cite{wang2004development} Partial charges for the carbonate ion were taken from restrained electrostatic potential (RESP) fitting reported therein and rescaled by a factor of 0.75 using the Electronic Continuum Correction (ECC) method.\cite{leontyev2009electronic,duboue2020practical} Lennard-Jones parameters were refit to reproduce the calcium–carbonate potential of mean force in water.

\subsection{Fluid phase calculations}
This section describes molecular dynamics simulations and RISM calculations of the bulk alkanes in the absence of the crystal surface, representing the homogeneous fluid phase.

\subsubsection{Molecular dynamics}
Cubic simulation boxes of approximately $\SI{50}{\angstrom}\times\SI{50}{\angstrom}\times\SI{50}{\angstrom}$, containing methane and ethane at bulk densities of \SI{9.3364e-3}{\angstrom^{-3}} and \SI{9.5616e-3}{\angstrom^{-3}}, respectively, were generated using \verb|AddToBox| command and parameterized with \verb|tleap|. 

Prior to molecular dynamics simulations, the system was energy-minimized using a combination of steepest descent and conjugate gradient methods. Following minimization, the system was heated from 0 to 320 K over 40 ps (20,000 steps, 2 fs timestep) under NVT conditions using a Langevin thermostat with a collision frequency of 2 ps$^{-1}$.

Production simulations were run under the same conditions for $4\times10^7$ steps (80 ns, 2 fs timestep). Nonbonded interactions were computed with a \SI{10}{\angstrom} cutoff under periodic boundary conditions, and coordinates were saved every 5000 steps.

\subsubsection{1D-RISM calculations}
Bulk-phase fluid properties were also calculated with extended 1D-RISM\cite{pettitt1982integral} (XRISM) as implemented in AmberTools \cite{luchko2010three, caseAmberTools2023}. The 1D-RISM equations were solved for 32,768 grid points with a grid spacing of \SI{0.025}{\angstrom}, a temperature of \SI{320}{\kelvin}, and a residual tolerance of $10^{-10}$. Convergence was acquired by using the modified direct inversion of the iterative subspace (MDIIS) method,\cite{kovalenko1999solution} using a maximum of 25 solution vectors, MDIIS step size of 0.3 and tolerance of $10^{-10}$. The number density for each alkane was set to match the values defined for MD calculations. 

\subsection{Interface calculations}
\subsubsection{Calcite slab preparation}
\begin{figure}
    \centering
	\includegraphics{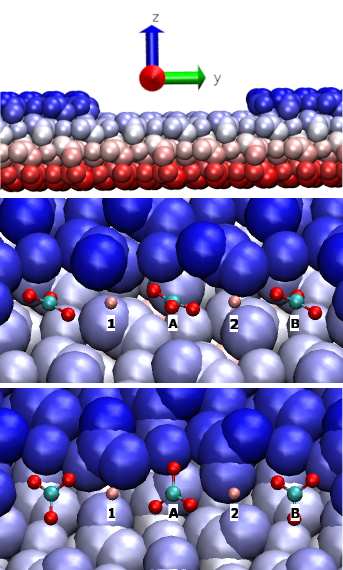}
	\caption{Side view (top) of the crystal slab containing the step obtuse (left) and acute (right) defects, the crystal is rendered as van der Waals spheres with a blue-white-red gradient indicating decreasing z-coordinates. Calcium (1 and 2) and carbonate labels (A and B) are discussed in the text  for  acute (center) and obtuse (bottom) steps. Atoms color as calcium: pink; carbon: cyan; oxygen: red.}\label{fig:step-labels}
\end{figure}

The slab was constructed from a triclinic unit cell containing eight CaCO\textsubscript{3} pairs in the same plane (see structure in the Supporting Information) and unit cell parameters $a=b=\SI{12.964}{\angstrom}$, $c=\SI{6.427}{\angstrom}$, $\alpha=78.98$\textdegree, $\beta=36.85$\textdegree~and $\gamma=101.93$\textdegree. The unit cell was replicated four times along each vector, resulting in a slab containing 512 CaCO\textsubscript{3} pairs. The dimensions of the periodic simulation box were set to \SI{51.856}{\angstrom}, \SI{51.856}{\angstrom}, \SI{80}{\angstrom}, 90.00\textdegree, 90.00\textdegree and 101.93\textdegree, ensuring space to accommodate the fluid during the simulations. A second system with step defects was acquired by removing half of the top layer along the $y$-direction. Each step contains two distinct calcium and carbonate sites, labeled according to Figure \ref{fig:step-labels}.

\subsubsection{Molecular dynamics}
For the MD calculations, each calculation was done with three replicas. The simulation box was filled with 1600 molecules of each alkane, which gave approximately the same density in the bulk region as for the fluid-phase calculations. The simulations at the interface were carried out using the same settings as for the bulk system. The simulation length of all systems was \SI{80}{\nano\second}.

As previously reported in the literature,\cite{kristensen2004modeling,nygren1998incorporation} rotations in some of the carbonates were observed in the step defects during simulations. Therefore, we carried out two types of simulations. In the first type, harmonic restraints (\SI{100}{\kcal\per\mol\per\angstrom^2}) were applied to the crystal atoms to prevent rotations and preserve the crystallographic structure. In the second type, unrestrained molecular dynamics simulations, in which only a single atom was fixed to prevent rigid-body translation, were used to assess the effect of the carbonate group rotation on the density profile of the alkanes near the step edges.

\subsubsection{3D-RISM calculations}

The density distributions of the fluid around calcite were acquired by solving the 3D-RISM equations \cite{luchko2010three, caseAmberTools2023}, using the results from 1D-RISM calculations to describe the solvent and the same force field as the MD simulations for the calcite crystal. In this step, the bulk densities for the 1D-RISM calculations were set to match the average number density in the bulk region in the MD fluid-surface simulations. The values acquired are presented in Table \ref{tab:bulk_densities}. A grid spacing of \SI{0.25}{\angstrom}, a maximum of 5 solution vectors, and a step size of 0.5 were used for MDIIS. The residual tolerance was set to $10^{-5}$, and periodic boundary conditions were employed.\cite{darden1993particle, grayIntegralEquationModels2022} To account for carbonate rotations and thermal vibrations within the 3D-RISM framework, calculations were also carried out for the stepped surfaces considering 81 different crystal configurations, sampled at regular intervals from a single MD trajectory, and the results averaged.

\begin{table}[h]
\centering
\caption{Bulk density values (\SI{}{\angstrom^{-3}}) for methane and ethane, acquired from MD simulations considering z coordinate interval $40 < z < 50$ \AA.}
\label{tab:bulk_densities}
\begin{tabular}{llll}
\hline
Alkane & Surface type & Restrained & $\rho_\text{bulk}$ \\
\hline
methane & Flat & Yes & $9.3364\times10^{-3}$ \\
methane & Step & Yes & $9.1179\times10^{-3}$ \\
methane & Step & No  & $9.0686\times10^{-3}$ \\
ethane  & Flat & Yes & $9.5616\times10^{-3}$ \\
ethane  & Step & Yes & $9.3594\times10^{-3}$ \\
ethane  & Step & No  & $9.3131\times10^{-3}$ \\
\hline
\end{tabular}
\end{table}

\section{Excess number calculation}
The excess adsorption per unit surface area was calculated according to

\begin{equation}\label{eq:excess}
\Gamma = \frac{1}{A} \int_V \left[ \rho(\mathbf{r}) - \rho_b \right] dV,
\end{equation}
where $A$ is the surface area, $\rho_b$ is the bulk fluid density, $V$ is the volume, and $\rho(\mathbf{r})$ is the local number density obtained either from MD simulations or from 3D-RISM calculations. 

The integration region was defined using a spatial mask constructed from the surface geometry. The entire $x$-$y$ domain was considered, while the lower $z$ limit was defined by the solvent-accessible surface. The upper limit was set to \SI{45}{\angstrom}, sufficiently far from the surface that the fluid density reaches its bulk value. Thus, the integration accounts for the excess number of particles within the solvent-accessible region above the surface.

For comparison between the 3D-RISM and MD results, equation (\ref{eq:excess}) can be expressed in terms of the total number of particles within the integration region as

\begin{equation}\label{eq}
\Gamma = \frac{1}{A}\left(N-\rho_b V_{\mathrm{SAV}}\right),
\end{equation}
where $N$ is the total number of particles within the integrated region and $V_{\mathrm{SAV}}$ is the corresponding solvent-accessible volume (SAV). This formulation allows the excess adsorption to be evaluated consistently from both the density distributions obtained from MD and 3D-RISM. 

To accurately account for the surface geometry, the solvent-accessible area at the crystal surface was determined at each MD frame using a probe with radius of \SI{0.85}{\angstrom}. The excess adsorption was then calculated for each frame, and the values were averaged first within each replica and then across replicas. The same procedure was applied to the 3D-RISM calculations performed on multiple surface configurations.

\section{Results}

\subsection{Fluid phase}
To establish a reliable reference for the liquid-phase structure, we first assess how accurately 1D-RISM reproduces intermolecular correlations in bulk methane and ethane. In particular, the radial distribution functions obtained from 1D-RISM are compared against molecular dynamics (MD) simulations.

\begin{figure*}
	\includegraphics{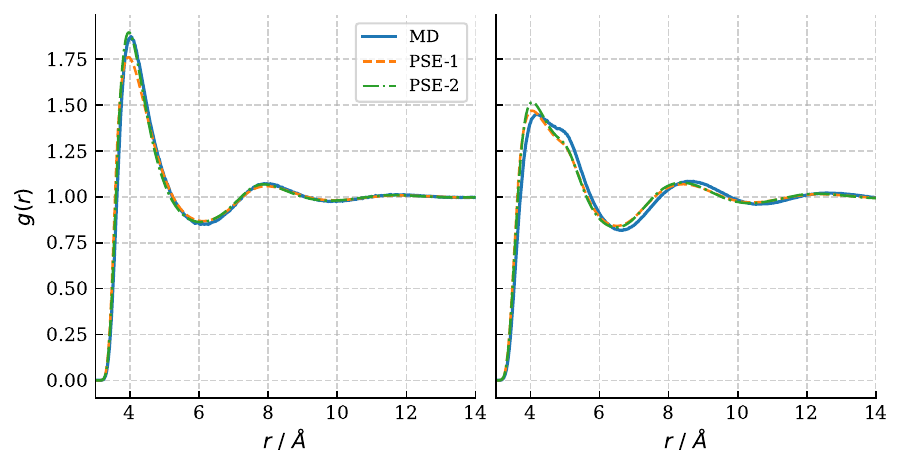}
	\caption{Radial distribution functions calculated from MD simulation and from 1D-RISM for (left) methane, with $\rho_\text{bulk} = \SI{9.3364e-3}{\angstrom^{-3}}$, and (right) ethane, with $\rho_\text{bulk} = \SI{9.5616e-3}{\angstrom^{-3}}$.}\label{fig:met-et-bulk}
\end{figure*}

Initially, the 1D-RISM pair correlation functions for the bulk fluids, using the PSE-1 and PSE-2 closure relations, were compared against MD results, obtained using the \verb|cpptraj|\cite{roe2013ptraj} program. The radial distribution functions from MD and 1D-RISM with PSE-1 and PSE-2 closures for methane and ethane are presented in Figure \ref{fig:met-et-bulk}. 

For methane, all three methods have a first peak near \SI{4}{\angstrom} and a second lower intensity peak near \SI{8}{\angstrom}, which correspond to the first and second  solvation shells. Excellent agreement is observed for PSE-2, while PSE-1 slightly underestimates the height of the first peak. 

For ethane, both MD and RISM predict the existence of two merged peaks at distances near \SI{4}{\angstrom} and \SI{5}{\angstrom}, resulting in a broad peak with a shoulder. This feature arises from the molecular structure of ethane. Since the RDF is calculated with respect to a single carbon site, the two carbon sites of a neighboring ethane molecule can contribute at different distances from the reference carbon, depending on the orientation of the molecule. The overlap of these contributions results in a first solvation-shell peak with a shoulder. The second solvation shell for ethane occurs at approximately \SI{8.5}{\angstrom}.

The primary difference between the two closures is that PSE-1 captures the maximum of the first peak well but underestimates the shoulder, while PSE-2  slightly overestimates the first peak maximum. As the PSE-2 closure best matches the overall results from MD simulations for methane and ethane, it is used for the subsequent 3D-RISM calculations in the presence of the crystal surface.

\subsection{Calcite-methane adsorption}
It is of particular interest whether methane exhibits preferential adsorption above specific surface sites, such as calcium or carbonate groups, and whether 3D-RISM can reproduce the same spatial organization observed in molecular dynamics simulations. To answer these questions, we analyzed the 3D number density distribution of methane in the presence of the flat calcite surface using MD and 3D-RISM-PSE-2. 

\begin{figure}[ht]
    \centering
    \includegraphics{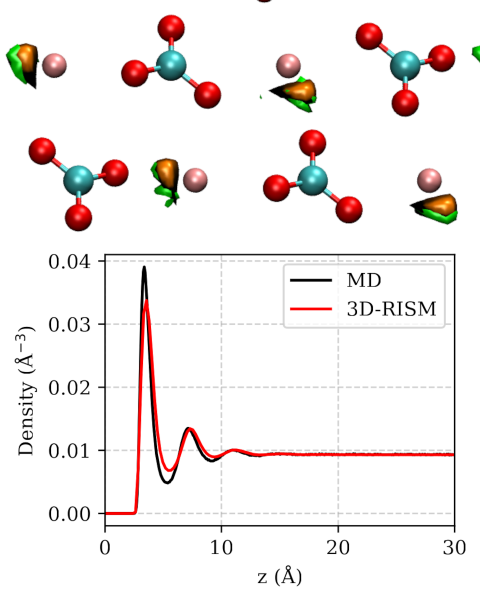}
	\caption{(Top) Methane number density profile isosurfaces near the maximum value (first peak) calculated from molecular dynamics (green isosurface, \SI{0.18}{\angstrom^{-3}}) and 3D-RISM (orange isosurface, \SI{0.097}{\angstrom^{-3}}). The view is normal to the calcite flat surface. Calcium ions are represented in pink, carbon in cyan and oxygen in red. (Bottom) Number density profiles for methane acquired from 3D-RISM and molecular dynamics averaged over the $x$-$y$ plane for different $z$ distance from the calcite plane.}\label{fig:methane-flat}
\end{figure}

Both methods predict elevated densities in close proximity to the calcium atoms, with a slight displacement toward the carbonate oxygen atoms (Figure \ref{fig:methane-flat}, top). The high-density regions correspond to minima in the potential of mean force (PMF), indicating energetically favorable local environments for methane. Furthermore, it demonstrates that 3D-RISM is able to capture the interactions between the methane and calcite with the same level of detail as MD simulations. 

The density profile along the $z$ direction shows a multilayer adsorption, with three clear peaks at approximately 3.5, 7.2 and \SI{11.2}{\angstrom} (Figure \ref{fig:methane-flat}, bottom). Although 3D-RISM underestimates the first peak and slightly overestimates the density within the depletion regions, it reproduces the positions and magnitudes of the subsequent peaks with good accuracy.

\begin{figure}[ht]
	\includegraphics{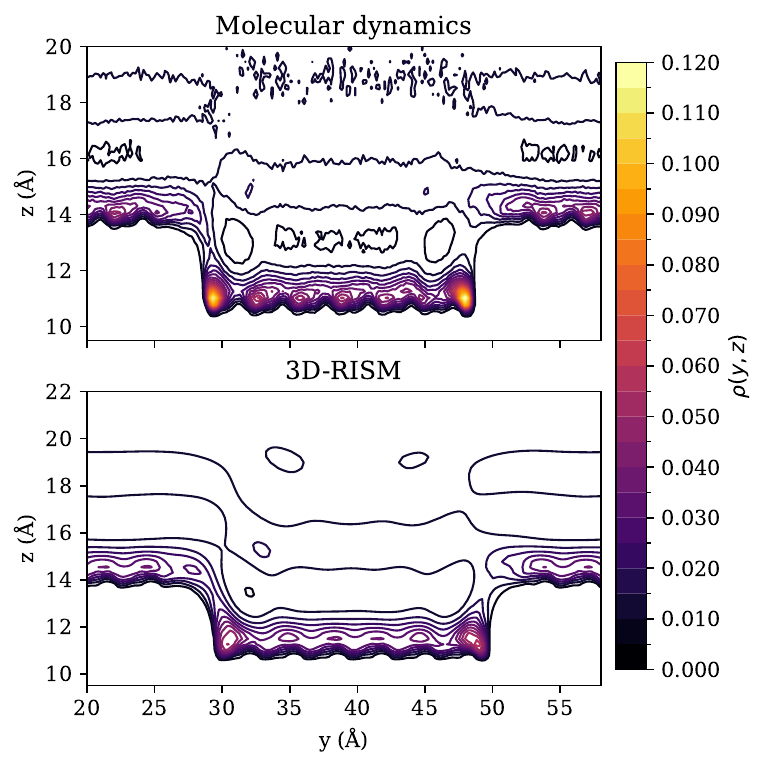}
	\caption{Contour plot of the density profile for methane for calcite with a step defect. The density is averaged along the $x$ direction. The obtuse angle is on the left and acute angle on the right. The maximum value in the color bar corresponds to the maximum density value acquired from MD simulations.}\label{fig:methane-rism-md-2d}
\end{figure}

When we consider a calcite crystal with a step defect, both methods predict a higher density region at the site of the defects, compared to the flat surface. This high density indicates higher affinity between the methane and calcite at the defect sites than the flat regions. Conversely, the first layer density is relatively low at the shoulders of the defects. Because methane and calcite interact only through the Lennard-Jones potential, this suggests that methane becomes sterically trapped in these crevices. 

3D-RISM again captures the distribution of methane, including the multilayer density and changes in density around the defect site. However, as seen in the flat surface calculations, 3D-RISM underestimates the highest peak intensity, with maximum densities only half that predicted by MD. This is primarily due to the closure approximation and could be mitigated by using other closure relations.\cite{luchko2010three, joungSimpleElectrolyteSolutions2013, giambasuIonCountingExplicitSolvent2014, stumpeCalculationLocalWater2011}

\begin{figure}
    \centering
	\includegraphics{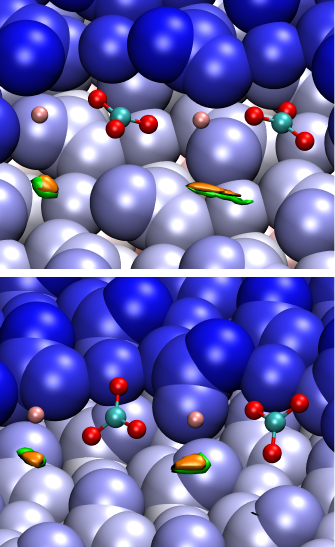}
	\caption{Methane density profiles at acute (top) and obtuse (bottom) step edges. The crystal representation is the same as in Figure \ref{fig:step-labels}. Density profiles for 3D-RISM are displayed in orange (isosurface of \SI{0.15}{\angstrom^{-3}}) and for MD in green (isosurface of \SI{0.30}{\angstrom^{-3}}).}\label{fig:methane-step}
\end{figure}

The effect of steric trapping can be seen by comparing the three-dimensional density profiles at the obtuse and acute angles (Figure \ref{fig:methane-step}). The elongated  distribution near the calcium site 2 in the acute defect indicates that methane is less localized. In contrast, the methane densities are all sharply peaked for the obtuse defect.

\subsection{Calcite-ethane adsorption}

Repeating the MD and 3D-RISM calculations with ethane presents a similar picture. Again, both methods predict elevated densities in regions located nearly directly above the calcium atoms, with a slight displacement toward the carbonate oxygen atoms (Figure SI2, top). There is little difference between the isosurfaces obtained for methane and ethane. Similarly, the density profiles show multilayer adsorption. The first peak for ethane is slightly lower than that for methane, whereas the second peak is more pronounced. Beyond the second layer, the ethane profile also exhibits weak oscillations extending farther from the surface, with four discernible peaks at approximately 3.5, 7.2, 11.5, and \SI{15.5}{\angstrom}, compared with only three peaks for methane. Overall, however, the density profiles of the two molecules are very similar, indicating comparable multilayer ordering near the surface.

When considering the step defect, we observe similar results for ethane as for methane. Ethane also has a higher density profile along the crevices of the step defects and has a relatively low density around the shoulders (Figure SI3). The distribution of ethane at the acute and obtuse angle defects (Figure SI4) is almost identical to that for methane (Figure \ref{fig:methane-rism-md-2d}). Both MD and 3D-RISM  predict lower maxima for the density peaks for ethane compared to methane. In addition, the difference in peak heights between MD and 3D-RISM is smaller for ethane.

While the differences between methane and ethane are subtle, we find that both methodologies predict a lower affinity of ethane compared with methane at approximately the same bulk number density. Furthermore, both methods show the accumulation along the step defects and differentiate between the obtuse and acute angles.

\subsection{MD: restrained vs unrestrained atoms}
Previous studies have shown that carbonate ions in the calcite step defects can change orientation when the structure is allowed to relax.\cite{kristensen2004modeling,nygren1998incorporation} To account for how this may change the density distributions of methane and ethane, molecular dynamics simulations were also performed without harmonic restraints on the crystal atoms. 

For the flat-surface simulations, the $z$-averaged density profiles of methane and ethane (Figure SI5 in Supporting Information) from MD simulations are almost the same as for the restrained MD simulations (Figure \ref{fig:methane-flat} and SI2). A previous study analyzing the flat surface in vacuum conditions and in contact with water\cite{demichelis2025reconstruction} concluded that flat surface reconstruction, {\it i.e.}, carbonate rotations, can occur in vacuum, but the preferred structure in the presence of water is the unreconstructed one. For the unrestrained simulations conducted in this work, carbonate was not observed to rotate on the flat surface.

\begin{figure*}
	\includegraphics{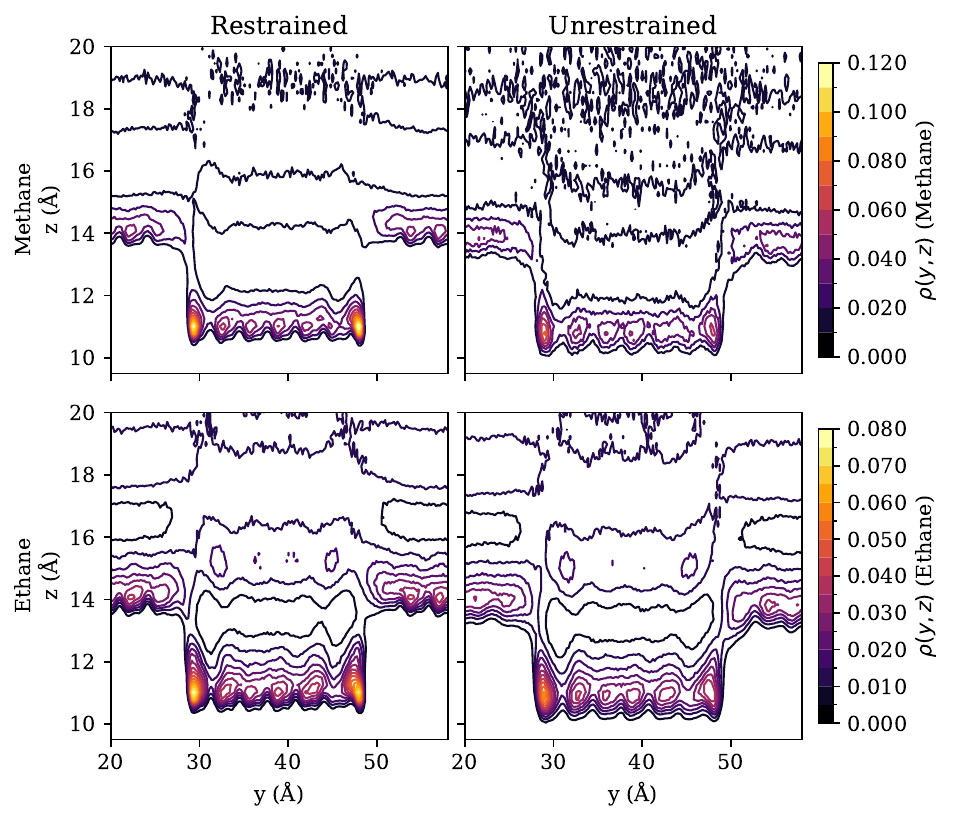}
	\caption{MD simulations of restrained and unrestrained two-dimensional density profile for methane and ethane averaged along the step edges. The color bar for each row has the maximum value defined based on the highest density from restrained simulations.}\label{fig:md1_md2_2d_compare}
\end{figure*}

\begin{figure}
    \centering
	\includegraphics{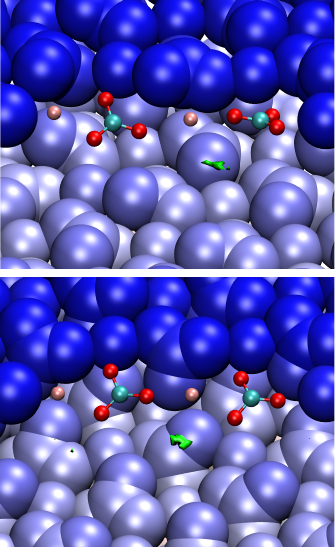}
	\caption{Density profile for methane at acute (top) and obtuse (bottom) steps acquired from unrestrained simulation: methane density is displayed in green at an isovalue of \SI{0.25}{\angstrom^{-3}} for both steps.}\label{fig:unrestrained-step}
\end{figure}

For the step defects, densities are reduced in both the acute and obtuse steps, with the acute step showing a less pronounced density peak, Figure \ref{fig:md1_md2_2d_compare}. This reduction arises from rotations and vibrations of some carbonate ions at the step edges, see Figures \ref{fig:unrestrained-step} and SI6. Comparing with Figure \ref{fig:methane-step}, it is observed that at the acute step, carbonate site A rotates while B retains its original orientation. At the obtuse step, both A and B rotate, eliminating the orientational difference between them. Similar structural changes were reported by Kristensen {\it et al.} \cite{kristensen2004modeling} in classical, static lattice simulations of the crystal under vacuum conditions.

The surface reconstruction observed at the step edges is persistent throughout the simulation. Figure SI6 shows the superposition of the step-site atomic positions sampled along the trajectory. The atoms are observed to fluctuate around the reconstructed configurations, with no evidence of returning to the original crystallographic positions. This indicates that the reoriented carbonate groups correspond to a stable relaxed surface structure rather than transient fluctuations away from the initial crystal geometry.

These changes in carbonate orientation lead to a significant reduction in the density peaks near sites 1 and 2 of the acute step and site 1 of the obtuse step. However, despite this decrease relative to the unrelaxed crystallographic structure, both methane and ethane still exhibit higher affinity for the step defects than for the flat regions. An energy minimization was carried out using 3D-RISM density profile as the solvent model and the carbonate rotations in the minimized structure also match those observed in the unrestrained MD simulations, see Supporting Information.

\subsection{Excess number of adsorbed particles}
To quantify the adsorption behavior at flat and stepped surfaces, we computed the excess number of adsorbed particles for both methane and ethane. This allows for a direct comparison between the 3D-RISM and MD results, and provides a scalar measure of how the introduction of step defects affects adsorption.

For the unrelaxed surfaces (Table \ref{tab:excess_methane_ethane}), the introduction of steps results in an increase in the calculated excess adsorption for both methane and ethane, although the magnitude of this change differs between the two methods. The MD results indicate a relatively modest change in the total excess adsorption upon introducing the steps. For methane, both methods predict an increase in excess adsorption upon going from the flat to the stepped surface. For ethane, this increase is more pronounced in 3D-RISM, whereas the corresponding difference in MD is small.

\begin{table}[h]
\centering
\caption{Excess adsorption per nm\textsuperscript{2} for methane and ethane on flat and step unrelaxed surfaces.}
\label{tab:excess_methane_ethane}
\begin{tabular}{llll}
\hline
Method & Surface type & $\Gamma_{\mathrm{methane}}$ & $\Gamma_{\mathrm{ethane}}$ \\
\hline
RISM & Flat & $2.421$ & $1.777$ \\
RISM & Step & $2.536$ & $1.973$ \\
MD   & Flat & $1.881 \pm 3\times10^{-3}$ & $1.168 \pm 3\times10^{-3}$ \\
MD   & Step & $2.017 \pm 1.2\times10^{-2}$ & $1.180 \pm 1.3\times10^{-2}$ \\
\hline
\end{tabular}
\end{table}

For the unrestrained stepped surface (Table \ref{tab:excess_methane_ethane_2}), the MD excess adsorption values remain comparable to those obtained for the restrained configuration for both molecules, suggesting that the carbonate rotations observed upon surface relaxation have a limited effect on the overall integrated adsorption. Nevertheless, the density distributions indicate that these structural changes locally modify the density peaks at specific step sites. In contrast, the 3D-RISM values averaged over 81 surface configurations are higher than those obtained for the single restrained configuration.

\begin{table}[h]
\centering
\caption{Excess adsorption per nm\textsuperscript{2} for methane and ethane on stepped surface. The averages for 3D-RISM were acquired using 81 surface configurations, while the MD values were averaged over three independent unrestrained replicas.}
\label{tab:excess_methane_ethane_2}
\begin{tabular}{lll}
\hline
Method & $\Gamma_{\mathrm{methane}}$ & $\Gamma_{\mathrm{ethane}}$ \\
\hline
RISM & $2.804 \pm 1.1\times10^{-2}$ & $2.318 \pm 1.3\times10^{-2}$ \\
MD   & $1.965 \pm 1.9\times10^{-2}$ & $1.138 \pm 7\times10^{-3}$ \\
\hline
\end{tabular}
\end{table}

Across both tables, 3D-RISM consistently predicts larger excess adsorption values than MD, while methane exhibits a higher excess adsorption than ethane for both methods. Thus, although the two approaches yield quantitatively different integrated excess adsorption values, they provide a consistent qualitative description of adsorption at the stepped surface.

The quantitative difference between the 3D-RISM and MD excess adsorption can be further examined through the three-dimensional density distributions. As shown in the contour profiles (Figures \ref{fig:methane-rism-md-2d}, SI3 and \ref{fig:md1_md2_2d_compare}) and density slices (Figures SI8--SI15), due to the closure approximation, 3D-RISM produces lower intensities for the highest-density peaks at the step sites and broader density distributions. Despite these quantitative differences, both methods identify the same main adsorption features, including preferential accumulation at the step defect sites and the asymmetry between the obtuse and acute geometries. Greater quantitative agreement of 3D-RISM predictions with MD may be achieved by improving the closure relation\cite{carvalho2021thermodynamic,tsednee2019closure} or developing an optimized set of potential parameters specific to the RISM framework.\cite{silva2026development,roy2023evaluating} 

It is worth noting that the stepped surfaces were generated by removing atoms only from the topmost layer, such that the resulting structural perturbation represents a relatively small fraction of the total surface area. Thus, calculations considering extended step faces, where the step exposes a distinct crystallographic plane, would provide a more realistic representation of the step geometries observed experimentally\cite{stack2010growth}.

\section{Conclusions}

In this study, we calculated the density distribution of methane and ethane at the surface of calcite crystals using MD and 3D-RISM. Both flat and step defect structures were considered. In addition, we simulated the effect of positional restraints on the distribution of both methane and ethane.

While 3D-RISM slightly underestimates density peak intensities and overestimates the excess number calculations, it successfully reproduces the trends observed in MD simulations for both methane and ethane. Across all systems, the adsorption is enhanced at the step defects, highlighting the critical role of surface heterogeneity in fluid-mineral interactions. Overall, the 3D-RISM density profile predictions were in good semi-quantitative agreement with the MD simulations for both the flat surface and the stepped surface, capturing the details of the overall distribution, though underestimating the highest density regions. This agreement demonstrates the reliability of 3D-RISM for modeling alkane adsorption on calcite surfaces. For flat surfaces, both restrained and unrestrained MD simulations yielded similar density distributions. In contrast, stepped surfaces underwent structural relaxation upon energy minimization, characterized by reorientation of carbonate groups at the step edges, consistent with previous studies.\cite{kristensen2004modeling,nygren1998incorporation} These structural changes affected the local density profiles near the defect, by reducing the density peak height next to specific step sites. The same behavior was obtained by averaging 3D-RISM results on different unrestrained surface conformations.

These results demonstrate that 3D-RISM is a computationally efficient tool for screening adsorption across a range of surfaces and conditions, providing insights semi-quantitatively comparable to MD simulations at a significantly lower computational cost. Both methods were run on 64 processor cores. Each MD production simulation required approximately 40–65 hours of wall time (2600–4200 core-hours). In contrast, each 3D-RISM calculation was completed in approximately 20 seconds (about 0.4 core-hours). In both cases, care must be taken when treating surface defects, as structural relaxation can strongly influence adsorption behavior. In such cases, short MD simulations can be used to generate conformations for subsequent 3D-RISM calculations.

Overall, we find that structural defects play a crucial role in methane and ethane adsorption on calcite surfaces. Accurately accounting for these features is essential for realistic predictions of adsorption properties, with implications for improving molecular-level strategies in enhanced oil recovery from complex geological systems and in the development of new adsorbent materials.

\section*{Acknowledgements}
The authors thank Prof.~Paolo Raiteri for kindly providing the calcite unit cell structure. This material is based upon work supported by the National Science Foundation (NSF) under grants CHE-2102668, OAC-2320718, and OAC-2320846.


\section*{Supporting information}
Details of the solvent force field preparation, calcite unit cell, ethane plots, unrestrained vs restrained simulation plots for the flat surface, crystal minimization with 3D-RISM, unrestrained simulation carbonate geometries and 2D density slices are given in the Supporting Information (PDF).

\bibliography{nalkanes.bib}

\newpage



  
  

\end{document}